\documentclass[11pt]{article}
\usepackage{moriond,epsfig,amssymb}
\usepackage{hyperref} 
\usepackage{amsmath,bm}
\def\be{\begin{equation}}
\def\ee{\end{equation}}
\def\bea{\begin{eqnarray}}
\def\eea{\end{eqnarray}}

\newcommand{\stau}{{\widetilde{\tau}}}

\newcommand{\mstau}{m_{\stau_1}}

\newcommand{\mgo}{{m_{\widetilde{g}}}}

\newcommand{\mG}{m_{\s G}}

\newcommand{\thest}{\theta_{\stau}}

\newcommand{\GEV}{\ensuremath{\,\textnormal{GeV}}}
\newcommand{\TEV}{\ensuremath{\,\textnormal{TeV}}}
\newcommand{\SEC}{\ensuremath{\,\textnormal{sec}}}

\newcommand{\s}[1]{\widetilde{#1}}

\newcommand{\Mp}{M_{\rm{Pl}}}

\newcommand{\TR}{T_{\rm{R}}}

\DeclareMathSymbol{\Gamma}{\mathalpha}{letters}{"00}
\DeclareMathSymbol{\Delta}{\mathalpha}{letters}{"01}
\DeclareMathSymbol{\Theta}{\mathalpha}{letters}{"02}
\DeclareMathSymbol{\Lambda}{\mathalpha}{letters}{"03}
\DeclareMathSymbol{\Xi}{\mathalpha}{letters}{"04}
\DeclareMathSymbol{\Pi}{\mathalpha}{letters}{"05}
\DeclareMathSymbol{\Sigma}{\mathalpha}{letters}{"06}
\DeclareMathSymbol{\Upsilon}{\mathalpha}{letters}{"07}
\DeclareMathSymbol{\Phi}{\mathalpha}{letters}{"08}
\DeclareMathSymbol{\Psi}{\mathalpha}{letters}{"09}
\DeclareMathSymbol{\Omega}{\mathalpha}{letters}{"0A}
\DeclareMathSymbol{\varGamma}{\mathalpha}{operators}{"00}
\DeclareMathSymbol{\varDelta}{\mathalpha}{operators}{"01}
\DeclareMathSymbol{\varTheta}{\mathalpha}{operators}{"02}
\DeclareMathSymbol{\varLambda}{\mathalpha}{operators}{"03}
\DeclareMathSymbol{\varXi}{\mathalpha}{operators}{"04}
\DeclareMathSymbol{\varPi}{\mathalpha}{operators}{"05}
\DeclareMathSymbol{\varSigma}{\mathalpha}{operators}{"06}
\DeclareMathSymbol{\varUpsilon}{\mathalpha}{operators}{"07}
\DeclareMathSymbol{\varPhi}{\mathalpha}{operators}{"08}
\DeclareMathSymbol{\varPsi}{\mathalpha}{operators}{"09}
\DeclareMathSymbol{\varOmega}{\mathalpha}{operators}{"0A}
\def\beq{\begin{equation}}
\def\eeq{\end{equation}}
\def\bea{\begin{eqnarray}}
\def\eea{\end{eqnarray}}
\def\bi{\begin{itemize}}
\def\ei{\end{itemize}}

\begin{document}
\vspace*{1.5cm}
\title{
Gravitino DM and high reheating temperatures after LHC 7/8
}

\author{JAN HEISIG}

\address{Institute for Theoretical Particle Physics and Cosmology,  \\ RWTH Aachen University, Germany
}

\maketitle\abstracts{
The presence of high reheating temperatures in the thermal history of the universe
challenges supersymmetric scenarios owing to the gravitino problem. We revise 
a general $R$-parity conserving gravitino dark matter scenario with a stau as the 
next-to-lightest superparticle being particularly constrained by searches for heavy 
stable charged particles at the LHC. Imposing a variety of experimental and theoretical 
constraints we show that points with $T_R>10^9 \GEV$ survive only in a very particular 
corner of parameter space.
}

\section{Introduction}

An attractive way to explain the baryon asymmetry in the universe is thermal 
leptogenesis\,\cite{Fukugita:1986hr}. For this mechanism to work the universe has 
to be heated up to temperatures $\TR\gtrsim10^{9}\GEV$\,\cite{Davidson:2002qv,Buchmuller:2004nz} 
in the post-inflationary phase of reheating. The recent observation of B-mode polarization of
the CMB reported by the BICEP2 collaboration\,\cite{Ade:2014xna} could well be explained
by GUT-scale inflation which in general is consistent with a reheating temperature in this 
ballpark.\footnote{As an example, assuming an inflaton mass of the order of 
$m_{\phi}\sim 10^{13}\GEV$ (which fits the BICEP2 data
in a simple chaotic inflation model with a quadratic potential)
and that the inflaton decays dominantly via Planck-suppressed dimension-five operators,
$ \Gamma_{\phi} \sim m_{\phi}^3/\Mp^2$, we obtain a reheating temperature in the ballpark of 
$\TR\sim\sqrt{\Gamma_{\phi} \Mp}\sim10^9\dots10^{10}\GEV$.}

Once we want to accommodate such a high reheating temperature in the early universe
supersymmetric scenarios potentially suffer from the gravitino problem\,\cite{Weinberg:1982zq}.
Massive gravitinos are typically not in thermal equilibrium in the post-inflationary universe
and the production due to thermal scattering in the hot bath renders their abundance 
proportional to the reheating temperature.\cite{Bolz:2000fu,Pradler:2006hh} 
Hence, a high reheating temperature leads to a high gravitino abundance. In a scenario with 
a neutralino as the lightest superparticle (LSP) late decays of the gravitino cause an additional 
energy injection during or after big bang nucleosynthesis (BBN) distorting the predictions for
the primordial abundances of light elements.\cite{Ellis:1984er} This imposes tight bounds on 
the maximal reheating temperature in this scenario and rules out thermal leptogenesis for a 
neutralino LSP and a gravitino mass of the order of the other sparticle masses.

One way to alleviate this problem is considering a gravitino LSP\,\cite{Fayet:1981sq}. 
In this case the reheating temperature is only constrained via the measured dark matter 
(DM) abundance. However, now the next-to-LSP (NLSP) can only decay into the gravitino 
via Planck-suppressed couplings (we assume $R$-parity conservation here) leading to a 
rather large NLSP life-time. In this case late NLSP decays can endanger successful BBN and 
further constrain the model. However, if the NLSP belongs to the MSSM it interacts at least 
weakly and so the NLSP abundance is determined from freeze-out. Hence, it is the MSSM 
parameters that govern the consistency with BBN constraints and not the reheating 
temperature. Further, the gravitino abundance shows a non-trivial dependence on the MSSM 
parameters through thermal and non-thermal contributions. As the MSSM parameters can 
in principle be measured---or so far at least be constrained---by LHC data it is natural to 
ask whether these measurements provide implications for the highest reheating 
temperatures that are still consistent with observations.

In this article we summarize our research\,\cite{Heisig:2013sva,Heisig:2013rya} 
which addresses this question considering the case of a stau NLSP\@. 
As in such a scenario the stau is stable on collider time-scales it provides 
a spectacular signature as a heavy stable charged particle (HSCP). The LHC is 
extremely sensitive to such a signal and thus wide implications can be derived. 
Other NLSP candidates are in general less constrained by the data.

In order to explore the SUSY parameter space we utilize a Monte Carlo scan
which we briefly summarize in section \ref{sec:mc}. 
In section \ref{sec:grav} we introduce the relevant mechanisms of gravitino production. 
We present our results in section \ref{sec:res}. Our study reveals the existence of 
parameter points that provide $\TR>10^9\GEV$ and survive all imposed constraints. 
All these points lie in a very particular corner of the SUSY parameter space. Those 
spectra feature a distinct signature at colliders that can be tested in the upcoming LHC 
run.

\section{Monte Carlo scan}\label{sec:mc}

In order to explore the implications of the LHC and the results of other experiments
on the highest reheating temperature we utilize a Monte Carlo scan over the SUSY 
parameter space.\cite{Heisig:2013rya} We do not restrict ourselves to any high scale 
model but vary the parameters freely at the TeV scale. 
We scanned over the 17-dimensional pMSSM parameter space with the following
input parameters and scan ranges:
\bea
-10^4\GEV \le &A_t &\le10^4\GEV \nonumber \\
-8000\GEV \le &A_b,\,A_\tau, \mu& \le8000\GEV \nonumber \\
1 \le & \tan\beta & \le 60 \nonumber \\
100\GEV \le &  m_A  & \le 4000\GEV \nonumber \\
200\GEV \le & \mstau & \le 2000\GEV \nonumber \\
\max(\mstau,700\GEV) \le & m_{\tilde{t}_1}, m_{\tilde{b}_1} & \le 5000\GEV \label{eq:scanranges}\\
0 < &  \thest, \theta_{\tilde{t}} & < \pi \nonumber \\
\mstau  \le & m_{\s L_{1,2}}, m_{\s e_{1,2}} & \le  4000\GEV \nonumber \\
\max(\mstau,1200\GEV)\le & \!\!m_{\s Q_{1,2}}\!\!
=  m_{\s u_{1,2}}\!= m_{\s d_{1,2}}\!\! & \le  8000\GEV \nonumber \\
\mstau  \le &  M_1, M_2 & \le  4000\GEV \nonumber \\
\max(\mstau,1000\GEV) \le & M_3 & \le  5000\GEV \nonumber 
\eea
The lighter stau was taken to be the NLSP and we required that at least one of the neutral 
$CP$-even Higgses, $m_h,m_H$, can be identified with the Higgs boson discovered at the 
LHC\,\cite{ATLAS-CONF-2013-014,CMS-PAS-HIG-13-005}:
$m_h\;\,\rm{or/and} \;\,m_H \in [123;128]\GEV$. We generated $10^6$ points obeying these
requirements. For each point the stau freeze-out abundance\,\cite{Belanger:2008sj} and a 
variety of observables were computed in order to apply experimental and theoretical bounds.
Limits on the sparticle mass spectrum were derived from a reinterpretation\,\cite{Heisig:2013rya} 
of the HSCP searches at the 7 and $8\TEV$ LHC by CMS\,\cite{CMS1305.0491}.
Further, we considered bounds from flavor\,\cite{HFAGbsgAug12,:2012ct}\,and 
precision\,\cite{Group:2012gb,Bechtle:2012jw}\,observables, collider searches 
for the MSSM Higgs sector\,\cite{Bechtle:2011sb} as well as theoretical constraints 
from charge or color breaking (CCB) minima\,\cite{Heisig:2013rya,Kitahara:2013lfa}.

For each point of the 17-dimensional pMSSM ten gravitino masses, $\mG$, were randomly 
chosen.\cite{Heisig:2013sva} The gravitino mass range was determined from the minimum 
stau life-time $\tau_{\stau_1}>10^4\SEC$ (for the lower edge) and from an upper bound on 
the stau life-time arising from searches for anomalously heavy hydrogen in deep sea water 
(for the upper edge).\cite{Heisig:2013sva} 
For the application of BBN bounds\,\cite{Jedamzik:2007qk,Jedamzik:2006xz} and bounds 
from diffuse gamma ray observations\,\cite{Kribs:1996ac} we computed the life-time and the 
hadronic branching ratios of the stau for each point.

\section{Production of gravitinos} \label{sec:grav}

On the one hand, gravitinos are produced thermally during reheating. 
The corresponding relic abundance reads\,\cite{Pradler:2006hh}
\beq
\Omega_{\s G}^{\text{th}}h^2 
\simeq  \sum_{i=1}^{3} c_i\, g_i^2(\TR) 
\left(1+\frac{M_i^2(\TR)}{3\mG^2}\right) 
\left(\frac{\mG}{100\GEV}\right) \left(\frac{\TR}{10^{10}\GEV}\right),
\label{eq:omegagrthermal}
\eeq
where $g_i$ and $M_i$ are the gauge couplings and the gaugino mass parameters, 
respectively, associated with the SM gauge groups $U(1)_Y$,  $SU(2)_{\text{L}}$, 
$SU(3)_{\text{c}}$ and $c_i$ are associated numerical factors of $\mathcal{O}(0.1)$. 
On the other hand, gravitinos are produced non-thermally from the decay of the NLSP 
after NLSP freeze-out. Due to the assumed $R$-parity conservation each stau eventually 
decays into a gravitino. Hence, the number density of staus before their decay, 
$n_{\stau_1}$, is equal to the number density of gravitinos after all staus have decayed, 
$n_{\s G}$, and thus
\beq
\label{eq:nonthprod}
\Omega_{\s G}^{\text{non-th}} h^2 
= \frac{\mG}{\mstau}\,\Omega_{\stau_1} h^2\simeq 3.7\times10^{-9} \,\mG Y\,,
\eeq
where we introduced the stau yield, $Y=n_{\stau_1}/s$, with $s$ being the entropy 
density. By demanding that the resulting total gravitino abundance matches the 
measured DM abundance, $\Omega_{\s G}^{\text{non-th}} h^2 +
\Omega^{\text{th}}_{\s G}h^2 =\Omega_\text{CDM}h^2$, we computed the required
abundance of thermally produced gravitinos, $\Omega^{\text{th}}_{\s G}h^2$. For 
$\Omega_\text{CDM}h^2$, we chose the best-fit value $\Omega_\text{CDM} h^2 = 
0.11889$\,\cite{Ade:2013zuv}. From \eqref{eq:omegagrthermal} we computed the 
reheating temperature, $\TR$, that provides $\Omega^{\text{th}}_{\s G}h^2$ for the 
given parameter point.

\section{Results and discussion} \label{sec:res}

\begin{figure}[h!]
\centering
\setlength{\unitlength}{1\textwidth}
\begin{picture}(0.5,0.35)
 \put(0.0,0){  
  \put(0.004,0.025){\includegraphics[scale=1.22]{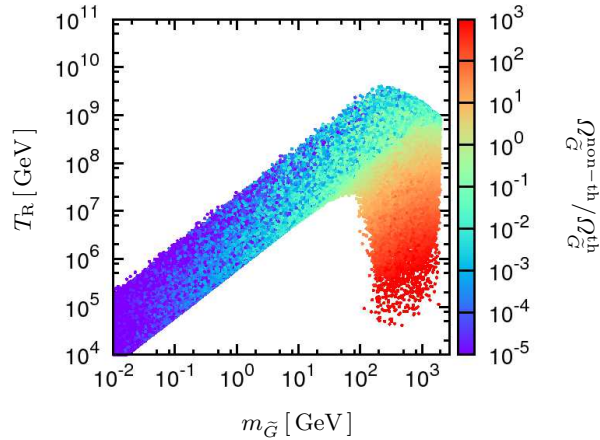}} 
  \put(0.19,0.0){\footnotesize $\mG\,[\GEV\,]$}
  \put(0.0,0.165){\rotatebox{90}{\footnotesize $\TR\,[\GEV\,]$}}
  \put(0.46,0.26){\rotatebox{-90}{\footnotesize $\Omega_{\s G}^{\rm{non-th}}/\Omega_{\s G}^{\rm{th}}$}}
  }
\end{picture}
\caption{Points of the $(17+1)$-dimensional pMSSM scan in the $\mG$-$\TR$ plane. 
Color code:~Ratio between the non-thermal and thermal contribution to the 
gravitino abundance, $\Omega_{\s G}^{\rm{non-th}}/\Omega_{\s G}^{\rm{th}}$.  \hfill \,
}
\label{fig:grav3}
\end{figure}

Figure \ref{fig:grav3} shows the ratio between the non-thermal and the thermal production 
of gravitinos. For small $\mG$ the non-thermal contribution is unimportant and the resulting 
reheating temperature grows linearly with the gravitino mass. Once the gravitino mass 
approaches the mass of the other superpartners we encounter two effects. 
First, according to \eqref{eq:omegagrthermal}, the linear growth of $\TR$ with $\mG$ turns 
into a decrease when approaching small mass splittings between the gravitino and the 
gaugino masses. This effect causes the points with the highest $\TR$ to lie around gravitino 
masses of a few hundred GeV. This number is a consequence of the chosen mass ranges 
for the gaugino mass parameters in our scan. The absolute maximum of $\TR$ reached
in our scan depends upon the lower limits of the scan ranges for the gaugino masses which 
are $200\GEV$ for $M_1$, $M_2$ and $1\TEV$ for $M_3$. It reaches $\TR\simeq
4\times10^9\GEV$ in accordance  with conservative limits found earlier\,\cite{Endo:2011uw}. 
%
As a second effect, once the gravitino mass approaches the mass of the stau NLSP\@, 
non-thermal contributions become important. Depending on the stau yield of a considered 
point the required reheating temperature is pushed down by a more or less significant amount. 
The points that still stay close to the upper edge of the populated band when $\mG$ approaches 
$\mstau$ tend to be those with very small yields. However,  we found points with yields 
$Y\gtrsim10^{-13}$ for $\TR\gtrsim10^9\GEV$. For these points the non-thermal contribution to 
the gravitino production is of the same order of magnitude as the thermal contribution and cannot 
be neglected.

\begin{figure}[t]
\centering
\setlength{\unitlength}{1\textwidth}
\begin{picture}(0.9,0.75)
 \put(-0.024,0.4){ 
  \put(-0.03,0.025){\includegraphics[scale=1.22]{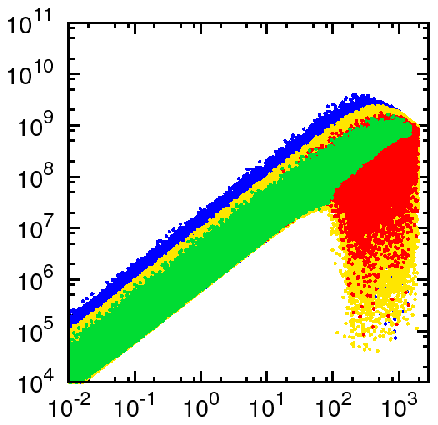}} 
  \put(0.19,0.0){\footnotesize $\mG\,[\GEV\,]$}
  \put(0.0,0.165){\rotatebox{90}{\footnotesize $\TR\,[\GEV\,]$}}
  }
 \put(0.46,0.4){  
  \put(-0.03,0.025){\includegraphics[scale=1.22]{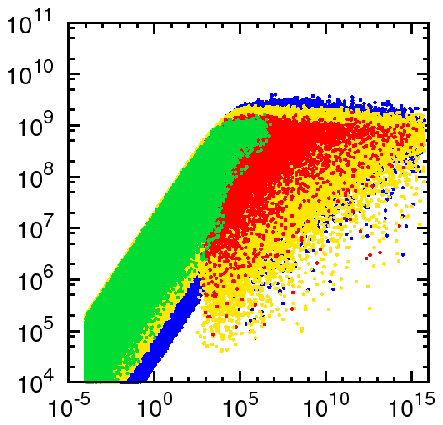} } 
  \put(0.205,0.0){\footnotesize $\tau_{\stau_1}\,[\,\sec\,]$}
  \put(0.0,0.165){\rotatebox{90}{\footnotesize $\TR\,[\GEV\,]$}}
  }
 \put(-0.024,0){ 
  \put(-0.03,0.025){\includegraphics[scale=1.22]{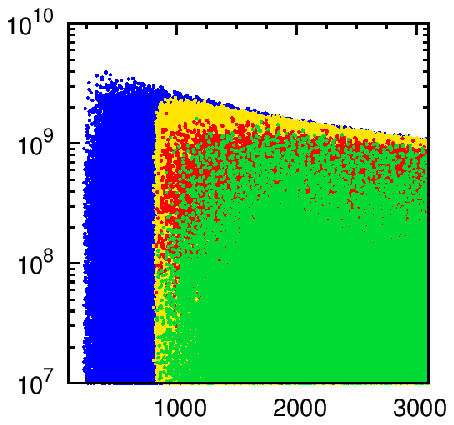}} 
  \put(0.19,0.0){\footnotesize $M_2\,[\GEV\,]$}
  \put(0.0,0.165){\rotatebox{90}{\footnotesize $\TR\,[\GEV\,]$}}
  }
 \put(0.46,0){  
  \put(-0.03,0.025){\includegraphics[scale=1.22]{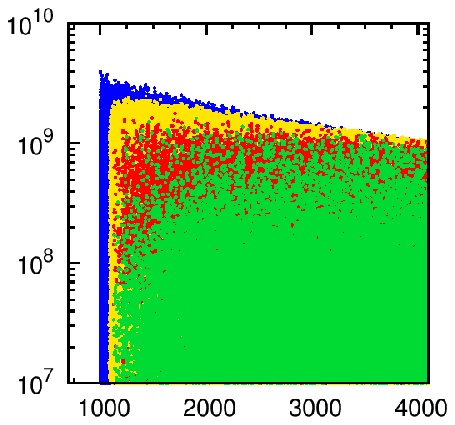} } 
  \put(0.205,0.0){\footnotesize $M_3\,[\GEV\,]$}
  \put(0.0,0.165){\rotatebox{90}{\footnotesize $\TR\,[\GEV\,]$}}
  }
\end{picture}
\caption{
Points of the $(17+1)$-dimensional pMSSM scan. The color code is chosen as follows. 
Blue:~Points passing no constraints. Yellow:~Points passing constraints from the HSCP 
search. Red:~Points additionally passing the constraints from flavor and precision 
observables, MSSM Higgs searches and CCB bounds. Green:~Points additionally passing 
the BBN bounds and bounds from the diffuse gamma ray spectrum. 
Upper left panel:~Reheating temperature $\TR$ against the gravitino mass $\mG$. 
Upper right panel:~Reheating temperature $\TR$ against the stau life-time $\tau_{\stau_1}$. 
Lower left panel:~Reheating temperature $\TR$ against the wino mass parameter $M_2$. 
Lower right panel:~Reheating temperature $\TR$ against the gluino mass parameter $M_3$.  \hfill \,
}
\label{fig:grav1}
\end{figure}

Figure \ref{fig:grav1} shows the effect of the bounds imposed on the $(17+1)$-dimensional 
parameter space. The blue points are rejected by the direct SUSY searches (i.e., the searches 
for HSCP). The yellow points are rejected by additional bounds from flavor and precision 
observables, MSSM Higgs searches or CCB bounds. The red points are rejected by the BBN 
bounds or the bounds from the diffuse gamma ray spectrum. The green points satisfy all 
constraints.
%
The searches for HSCP at the 7 and $8\TEV$ LHC impose very restrictive limits on the gluino 
and wino masses.\cite{Heisig:2013rya,Heisig:2012zq} In our scan we do not find allowed 
points with $\mgo<1.2\TEV$ or $M_2<800\GEV$\,\footnote{These limits can be understood 
as conservative limits on the individual parameters. Since we combine production channels 
for the derivation of exclusion limits in our scan, in general points with masses above these
limits can be excluded.} (see lower panels of figure \ref{fig:grav1}). 
Hence, these searches exclude all points with a reheating temperature above $\TR\simeq
2.3\times10^9\GEV$ (cf. blue versus yellow points). The very strong and robust limits on the 
SUSY masses are a particular feature of the stau NLSP scenario.\cite{Heisig:2013rya,Heisig:2012zq} 
For other choices of the NLSP these bounds can be considerably weaker, potentially leaving 
more room for $\TR\gtrsim10^9\GEV$. Note that the bound on $M_2$ is particularly important.
The gaugino masses in \eqref{eq:omegagrthermal} have to be evaluated at the scale $\TR$. 
Due to the faster running of $M_2$ with respect to $M_3$ up to $\TR$ the $SU(2)_{\text{L}}$ 
contribution can easily become dominant despite the smaller coupling.
Bounds from flavor and precision observables, MSSM Higgs searches and CCB vacua 
further reduce the parameter space leaving a maximal reheating temperature of slightly 
below $2\times10^9\GEV$ (cf. yellow versus red points) in our scan. The application of BBN 
bounds has the most significant effect in the region of large 
$\Omega_{\s G}^{\text{non-th}}/\Omega_{\s G}^{\text{th}}$ where $Y$ and $\mG$ (and therefore 
$\tau_{\stau_1}$) are large.

Despite the restrictive limits from HSCP searches in this scenario we find points which provide 
reheating temperatures $\TR>10^9\GEV$ and are consistent with all discussed bounds and 
with a Higgs mass of around $125\GEV$. All these points share very distinct features. 
First, these points feature gaugino masses not far above their respective lower limits
imposed by HSCP searches and a relatively heavy gravitino, $300\GEV<\mG<1.4\TEV$, 
in order to minimize \eqref{eq:omegagrthermal}. Second, BBN bounds and bounds from the
diffuse gamma ray spectrum disfavor very large life-times and do not allow for 
$\tau_{\stau_1}>10^7\SEC$ in our scan (see upper right panel of figure \ref{fig:grav1}). 
Hence, we encounter a separation between the gravitino mass and the stau mass of at least 
$200\GEV$. This separation coincides with the one between the gravitino mass and the 
gaugino masses in such a way that we find rather small mass splittings between the stau 
and the gauginos. This is most pronounced for $M_2$. As a consequence the strong bounds 
on $\mgo$ and $M_2$ also lift up the stau masses for points with $\TR>10^9\GEV$ in our 
scan, which we found to lie above $\mstau\simeq800\GEV$. Third, from BBN bounds those 
points are required to feature exceptionally small yields $Y<3\times10^{-14}$ being allowed 
only in region of parameter space where annihilation dominantly occurs via a resonant heavy 
Higgs in the $s$-channel, $m_A\simeq2\mstau$.\cite{Heisig:2013rya} For most points the 
dominant annihilation process is resonant stau-pair annihilation\,\cite{Pradler:2008qc}.
However, we also found a few points where resonant stop or EWino co-annihilation\,\cite{Heisig:2013rya} 
is the dominant process. Note that EWino co-annihilation via a resonant heavy Higgs requires 
no particularly large Higgs-sfermion couplings. Thus, the viability of these points does not rely 
on constraints from CCB vacua. As in this case the annihilation is driven by pair-annihilating 
EWinos it shows that similarly small yields could as well be achieved in a neutralino NLSP 
scenario.

\section{Conclusion}

In this article we examined the interplay between constraints on the SUSY parameter 
space and the highest possible reheating temperatures in a gravitino-stau scenario 
taking into account the thermal and non-thermal production of gravitinos. We found 
valid points with a reheating temperature high enough to allow for thermal leptogenesis, 
$\TR\gtrsim10^9\GEV$. These points are consistent with BBN bounds, flavor and 
precision bounds, theoretical bounds from vacuum stability, bounds from the HSCP 
searches at the 7 and 8\,TeV LHC as well as bounds from the MSSM Higgs searches 
and the requirement of providing a Higgs around 125\,GeV. In order to pass the BBN 
bounds all these points feature exceptionally small stau yields, $Y\lesssim10^{-14}$, 
that are only allowed in the resonant region, $m_A\simeq2\mstau$. In this region
annihilation dominantly takes place via the exchange of an $s$-channel heavy Higgs
either via resonant stau pair annihilation or resonant co-annihilating sparticles. 

\pagebreak

For most of the points with $\TR>10^9\GEV$ the dominant production mode at the 
$13/14\TEV$ LHC would be the production of EWinos or gluinos being relatively close 
in mass to the stau. Further, due to the resonant configuration, $m_A\simeq2\mstau$, 
resonant stau production via the $s$-channel heavy Higgs would be an important 
contribution. At the $13/14\TEV$ LHC this open window for high reheating temperatures 
can be tested. 

The gaugino masses $M_2$ and $M_3$ are of particular importance here. For other
NLSP candidates the respective mass limits can be much weaker than in the present 
case of a stau NLSP\@. For $M_2$ and $M_3$ close to the lower edges of our scan 
ranges, $M_2\gtrsim200\GEV$ and $M_3\gtrsim1\TEV$ we found a maximum reheating 
temperature of around $\TR\simeq4\times 10^9\GEV$. Provided an equally efficient 
annihilation of the NLSP candidate and similar constraints from BBN we expect 
temperatures around this value to be maximally allowed in a neutralino NLSP scenario.

\section*{Acknowledgements}

I would like to thank J\"orn Kersten, Boris Panes and Tania Robens for a fruitful 
collaboration and Valerie Domcke and Marco Drewes for very helpful discussions.
Further, I wish to thank the Moriond organizers for financial support. This work was 
partly supported by the German Research Foundation (DFG) via the Junior Research 
Group `SUSY Phenomenology' within the Collaborative Research Center 676 
`Particles, Strings and the Early Universe'.

\section*{References}

\addcontentsline{toc}{chapter}{References}
\providecommand{\href}[2]{#2}\begingroup\raggedright\endgroup

\end{document}